\documentstyle[preprint,aps,prl,psfig]{revtex}
\tightenlines
\begin{document}
\newcommand{\sinf}{\raisebox{-.7ex}{$\stackrel{<}{\sim}$}}
\newcommand{\ssup}{\raisebox{-.7ex}{$\stackrel{>}{\sim}$}}

\ifx\undefined\psfig\def\psfig#1{    }\else\fi
\ifpreprintsty\else
\twocolumn[\hsize\textwidth
\columnwidth\hsize\csname@twocolumnfalse\endcsname       \fi    \draft
\preprint{  }  \title  {Spin Coulomb drag in the two-dimensional electron
liquid}
\author  {Irene D'Amico}
\address{ Istituto Nazionale per la Fisica della Materia (INFM);\\
Institute for Scientific Interchange, via Settimio Severo 65, 
I-10133 Torino,
Italy}
\author{Giovanni Vignale}
\address{Department  of Physics,   University   of
Missouri, Columbia, Missouri 65211} 
\date{\today} \maketitle

\begin{abstract}
We calculate the spin-drag transresistivity $\rho_{\uparrow \downarrow}(T)$
in a
two-dimensional electron gas at temperature $T$ in the random phase 
approximation. In the low-temperature regime
we show that, at variance with the three-dimensional low-temperature result
[$\rho_{\uparrow\downarrow}(T) \sim T^2$],  the  spin transresistivity of
a two-dimensional
{\it spin unpolarized} electron gas  has the form
$\rho_{\uparrow\downarrow}(T) \sim  T^2
\ln T$.   In the spin-polarized case the familiar form
$\rho_{\uparrow\downarrow}(T) =A T^2$ is recovered, but the constant of
proportionality
$A$ diverges logarithmically as the spin-polarization tends to zero.   In
the high-temperature
regime we obtain 
$\rho_{\uparrow \downarrow}(T) = -(\hbar / e^2) (\pi^2 Ry^* /k_B T)$
(where $Ry^*$
is the effective Rydberg energy)  {\it independent} of the density.  Again, this
differs from the
three-dimensional result, which has a logarithmic dependence on the density.
Two important differences between the spin-drag transresistivity and the
ordinary
Coulomb drag transresistivity  are pointed out:
(i) The $\ln T$ singularity  at low temperature is smaller, in the Coulomb
drag case, by a factor
$e^{-4 k_Fd}$ where  $k_F$ is the Fermi wave vector and $d$ is the
separation between the
layers.  (ii)   The collective mode contribution to the spin-drag
transresistivity is negligible at all
temperatures.
Moreover the spin drag effect is, for comparable parameters,
 larger than the ordinary Coulomb drag effect.
\end{abstract}
\ifpreprintsty\else\vskip1pc]\fi \narrowtext
\section{Introduction}
The problem of transporting an electronic  spin
polarization from a place to
another by means of an electrical current  has attracted tremendous
interest in recent years,  both
in theoretical and experimental circles\cite{gen}.  A particularly  challenging
problem is that of injecting
spin-polarized electrons from a ferromagnet into an ordinary non-magnetic
semiconductor such
as $GaAs$.    This is by no means easy, due to the large impedance mismatch
between the
ferromagnetic injector and the semiconductor\cite{Schmidt}.  
However, the discovery that $GaAs$,
when doped with
magnetic $Mn$ impurities,  becomes ferromagnetic  with a Curie
temperature
$T_c$  as large as $110 K$ \cite{Ohno} has raised hopes of realizing useful
all-semiconductor
spin-electronic devices in a near future.

In dealing with spin-polarized currents one must be prepared to treat a
situation in which
electrons of opposite spin orientations travel, on the average, with
different speeds.  For
example, the up-spin electrons injected from a ferromagnet into a
semiconductor might be
drifting in one direction, while the down spin electrons remain on the
average stationary.
Whenever such a situation occurs, the existence of Coulomb correlation
between  up- and
down-spins becomes important, because it is a source of {\it friction}
between the two
components.    In our previous works \cite{sdrag,EPL,long} we examined the
effect of
up-down spin correlations on spin polarized transport in a
three-dimensional electron gas, and we
introduced the  concept of spin Coulomb drag (SCD)\cite{sdrag}: due to
Coulomb scattering,
momentum is  transferred  between the spin populations,  tending to
equalize their average drift
velocities. This process represents an actual drag exerted by the slower
population on the
faster  and, in the absence of an external driving field,  leads to a decay
of the spin current even
in the absence of extrinsic impurities\cite{spin-flip}. Later, it was shown that the SCD
also limits the
diffusion of a spin-packet in a doped semiconductor (\cite{EPL}) adding to
the conventional
effect of impurity scattering.

The quantitative measure of the SCD is the   spin-drag
transresistivity $\rho_{\uparrow\downarrow} (T)$, which is defined as the
ratio of the gradient of
electrochemical potential for up-spin electron ($E_\uparrow$) to the
current of down-spin
electrons  ($j_\downarrow$) when the current of up-spin electrons is zero:
\begin{equation}
\label{transresistivity} E_\uparrow =  \rho_{\uparrow\downarrow} (T)
j_\downarrow,~~~~~~(j_\uparrow = 0). \end{equation}
In Ref. \cite{sdrag} we presented a calculation of $
\rho_{\uparrow\downarrow}
(T)$  in a  pure three dimensional electron gas in the Random Phase 
Approximation (RPA).
The calculation is analogous to that of the more familiar Coulomb drag
effect \cite{Rojo},
but, as we will show momentarily, some important differences exist
between the two
effects.

The ordinary Coulomb drag (CD) effect involves {\it two}
distinct
quasi-two-dimensional electron layers that are spatially separated by an
insulating barrier. By contrast, in  the SCD,   we are  dealing with a {\it
single} layer in
which the two carrier populations (distinguishable by the different spin)
share the same space and
interact with the same  impurities\cite{impurities}.
 It follows that the SCD can be  either a
two-dimensional (2D) or a
fully three-dimensional (3D) effect, depending on the dimension of the
space in which the
electrons move. We have already  shown that the 3D spin transresistivity
should be measurable in
metals\cite{sdrag} and becomes larger in semiconductors \cite{EPL,long},
where it can be of
the same order of magnitude as the usual Drude resistivity. In this paper
we  concentrate on the
2D case.    Due to the absence of spatial separation between the two spin
populations, the bare
interaction between parallel and antiparallel spin electrons coincides
 with the interaction between parallel spin electrons, and
its Fourier transform
is simply $v(q) = 2 \pi e^2/q \epsilon$ where $q$ is the wave vector in the
plane and $\epsilon$ is
the appropriate low-frequency dielectric constant\cite{phonons}.  By
contrast, in the standard
CD case, the bare interaction between electrons in the two layers has the
2-dimensional Fourier
transform $v_{12}(q) = 2 \pi e^2 e^{-qd}/q \epsilon$ where $d$ is the
distance between the
layers. This ``innocent" difference has two important consequences.  First,
there is no
acoustic-plasmon-mediated interaction in SCD, whereas the effect of
acoustic plasmons on the
temperature dependence of the CD in separate layers is quite marked 
\cite{plasmon}.  Second, the
effective interaction between up and down spins in SCD remains sizable up
to  values of $q$
comparable to the Fermi momentum:  the presence of a significant
interaction strength  at $q=2
k_F$ is responsible for a characteristic logarithmic singularity in the
low-temperature
transresistivity.    

We provide analytical expressions  for
$\rho_{\uparrow \downarrow}$ in
both the low- and high-temperature limits.    At low temperature we derive
analytically the  $T^2 \ln T $
singularity;
at high temperature we show that $\rho_{\uparrow \downarrow}$
becomes
density-independent. Both these limits are significant  for realistic
  metallic and semiconductor parameters, 
respectively.
 We also draw a  comparison between the spin Coulomb
drag and the
ordinary Coulomb drag.  
It is hoped that the theoretical results presented
in this paper will
stimulate experimental work aimed at a direct observation of the SCD and a
comparison between
SCD and CD\cite{Flensberg}.

\section {Spin transresistivity in the 2-dimensional electron liquid}
 To leading order in the strength of the Coulomb interaction\cite{footnote},
 the 
spin-drag  transresistivity
is given by Eq. (18) of Ref. \cite{sdrag}, which, in the static case and
after few simple
rearrangements, can be written
as
 \begin{eqnarray}
\rho_{\uparrow\downarrow}(T) &=&
-\frac{\hbar}{e^2}\frac{\hbar }{n_\uparrow n_\downarrow k_B T }
\frac{C_D}{(2\pi)^{D+1}}\int_0^\infty{d q }
\frac{q^{D+1}}{D}
\cdot\nonumber
\\ &~&
\int_{0}^\infty d\omega 
\frac{Im Q_{0\uparrow}(q,\omega )
Im Q_{0\downarrow}(q,\omega )}{\left|\epsilon(q,\omega )\right|^2
\sinh^2(\hbar\omega /2k_BT)}.
\label{rho}
\end{eqnarray}
Here $k_B$ is the Boltzmann constant,
$D$ is the number of spatial dimensions and 
$C_D=1,~2\pi,~4\pi$ for $D=1,~2,~3$ respectively.
The dimensionless function $Q_{0 \sigma} (q,\omega)$ is defined as
\begin{equation}
Q_{0 \sigma} (q,\omega) \equiv
v(q) \chi_{0
\sigma}(q,\omega), 
\end{equation}
where $\chi_{0
\sigma}(q,\omega)$
 is the  non-interacting density-density response
function (the
``Lindhard function")   for spin $\sigma =
\uparrow$ or
$\downarrow$ and   $v(q)$  is the Fourier transform  of the Coulomb
interaction.  $\epsilon
(q,\omega) = 1 - Q_{0 \uparrow} (q,\omega)  - Q_{0 \downarrow} (q,\omega)$
is the RPA
dielectric function of the electron liquid, and  $n_\uparrow$ and
$n_\downarrow$ are the
densities of up- and down-spin electrons.  
  In both two and three dimensions
$ Im \chi_{0\sigma}$
can be expressed analytically. In particular, in two-dimensions,   starting from the
familiar expression
\begin{eqnarray}
Im \chi_{0\sigma}(q,\omega)&=&-\pi\int{d^2k\over(2\pi)^2}n_k
\left\{\delta (\hbar\omega+E_k-E_{k-q})\right. \nonumber \\ -
&~&\left.\delta (\hbar\omega-E_k+E_{k-q})\right\},
\end{eqnarray} with $E_k=\hbar^2 k^2/2m^*$ and
$n_k=1/[\exp((E_k-\mu)/k_B
T)+1]$
 ($m^*$ is the effective mass), we obtain
\begin{eqnarray}
Im Q_{0\sigma}(q,\omega) &=&
-{k_{F\sigma}\sqrt{\bar T_\sigma}\over q^2 a^*}
\big  [ F\left(e^{(\nu_{\sigma -}^2-\bar{\mu}_\sigma)/\bar T_\sigma}\right)
\nonumber \\ &-&
F\left(e^{(\nu_{\sigma +}^2-\bar{\mu}_\sigma)/\bar T_\sigma }\right) \big  ].
\label{Q2D}
\end{eqnarray}
Here $a^* = \hbar^2 \epsilon/m e^2$ is the effective Bohr
radius, $k_{F \sigma}$ is the Fermi
wave vector of $\sigma$-spin electrons,
$\bar
T_\sigma = k_BT/E_{F \sigma} $ and  $\bar \mu_\sigma = \mu_\sigma/E_{F
\sigma}$ are the
temperature and  chemical potential of   $\sigma$-spin electrons  expressed
in units of the
$\sigma$-spin Fermi energy $E_{F \sigma}$,  and
\begin{equation}  {\nu}_ {\sigma\pm}\equiv
{\omega \over{q} v_{F\sigma}}  \pm {q \over 2 k_{F \sigma}},
\end{equation}
where  $v_{F \sigma} = \hbar k_{F
\sigma}/m^*$ is the $\sigma$-spin 
Fermi velocity.

The function $F$ in Eq.~\ref{Q2D} is defined as
\begin{equation} \label{defF}
F(z)=\int_0^\infty {dx\over z e^{x^2}+1}.
\end{equation}

In Fig.~\ref{fig2} we present the  behavior of  $\rho_{\uparrow\downarrow} (T)$
calculated from Eq.~(\ref{rho}) and $D=2$ for three different values of the
carrier density with the values of the effective mass and dielectric
constant appropriate for
$InAs$  ($m^*=0.026m_e$, $m_e$ the free electron mass and $\epsilon = 13.6$).  As
pointed out in \cite{sdrag,long},  the spin-transresistivity increases with
decreasing densities.
As a function of temperature, $\rho_{\uparrow \downarrow}$ vanishes at
$T=0$,  reaches a
maximum about the Fermi temperature $T_F$, and decreases for large $T$.
As shown in  Fig.~\ref{fig2}, the overall
scale of the effect, at its largest, is set by  $\hbar /e^2 \simeq  4.1 k
\Omega$. This is quite a
large value and should definitely be observable.

We  now   derive and discuss the degenerate and non degenerate limits of
Eq.~(\ref{rho}).

\section {The degenerate limit}  In the limit  $T\ll T_F$ the calculation of
$\rho_{\uparrow \downarrow} (T)$ can be carried out  analytically.  The
temperature
dependence of $\rho_{\uparrow \downarrow}$, in this regime,  is  entirely
determined by the
denominator $\sinh^2( \hbar \omega/2k_BT)$, which restricts the integral in
Eq. (\ref{rho}) to
frequencies of the order of $k_BT$.    We can therefore neglect the
temperature dependence of
 $Im Q_{0\sigma}$ and replace this function  by its zero-temperature limit
\begin{eqnarray}
Im Q_{0\sigma}(q,\omega)&=&-{k_{F \sigma} \over q^2a^* }
\left[\Theta(1-{\nu}_{\sigma -}^2)\sqrt{1-{\nu}_{\sigma -}^2}
\right. \nonumber \\
&-&\left.\Theta(1-{\nu}_{\sigma +}^2)\sqrt{1-{\nu}_{\sigma +}^2}
\right].
\label{chi_T0}
\end{eqnarray}
The  behavior  of $Im Q_{0 \sigma}(q,\omega)$ for small $\omega$ depends
crucially on the
value of $q$.  As shown in the upper panel of 
Fig. 2, there are three distinct regions in which
$Im Q_{0 \sigma}(q,\omega)$ is different from zero:
\begin{itemize}
\item
(I) $q_{A\sigma}\equiv  k_{F\sigma}\left[-1+\sqrt{1+{2\omega \over k_{F\sigma}
v_{F\sigma}}}\right]
 < q < k_{F\sigma}\left[1-\sqrt{1-{2\omega \over k_{F\sigma}
v_{F\sigma}}}\right]
\equiv q_{B\sigma}$.
\begin{equation}
Im Q_{0 \sigma}(q,\omega) \simeq -{k_{F \sigma} \over q^2a^*} \sqrt{{\omega
\over k_{F
\sigma} v_{F \sigma}}}.\label{region1}
\end{equation}
\item
(II) $q_{B\sigma}\equiv  k_{F\sigma}\left[1-\sqrt{1-{2\omega \over k_{F\sigma}
v_{F\sigma}}}\right] < q <k_{F\sigma}\left[1+\sqrt{1-{2\omega \over k_{F\sigma}
v_{F\sigma}}}\right]\equiv q_{C\sigma}$
\begin{equation}
Im Q_{0 \sigma}(q,\omega) \simeq -{k_{F \sigma} \over  q^2a^*} { \omega /
(k_{F \sigma}
v_{F \sigma}) \over \sqrt{1 - (q/2 k_{F \sigma})^2}},
\label{region2}
\end{equation}
\item
(III) $q_{C\sigma}\equiv k_{F\sigma}\left[1+\sqrt{1-{2\omega \over k_{F\sigma}
v_{F\sigma}}}\right]< q <k_{F\sigma}\left[1+\sqrt{1+{2\omega \over k_{F\sigma}
v_{F\sigma}}}\right] \equiv q_{D\sigma}$.   \begin{equation}
Im Q_{0 \sigma}(q,\omega) \simeq -{k_{F \sigma} \over  q^2a^*} \sqrt{{\omega
\over k_{F
\sigma} v_{F \sigma}}+1 -{q^2\over 4 k_{F \sigma}^2}},\label{region3}
\end{equation}
\end{itemize}

Let us first consider  the paramagnetic case $n_\uparrow = n_\downarrow$,
$k_{F \uparrow} = k_{F \downarrow}$ etc..  In  region I  the product $ Im
Q_{0 \uparrow}
Im Q_{0 \downarrow}$ is proportional to $\omega$, while the integration
region in q space is a
shell of thickness proportional to $\omega^2$.  The net q-dependence of the
integrand (taking into account the fact that 
$\epsilon (q,0) \simeq 1 +  2 /(qa^*)$,
$|\epsilon(q,0)|^2 \sim  1/
 (qa^*)^2$)  is $\propto q$, giving an extra factor $\omega / v_F$.  Since
$\hbar\omega \sim k_B T$, the
contribution of this region  to $\rho_{\uparrow \downarrow}$ is at least of
order $T^4$ and will
be disregarded hereafter.

In  region II  the product $ Im Q_{0 \uparrow}
Im Q_{0 \downarrow}$ is proportional to $\omega^2$, while the integration
region in q space is a
shell of thickness $\sim 2 k_F$.  This would give a contribution of order
$T^2$ (as in 3
dimensions) were it not for the square root divergence of  $Im Q_{0
\sigma}$ when $q$
approaches $2 k_F$  [see Eq. (\ref{region2})].  Due to the ``piling-up" of
{\it two} square-root singularities, one gets an additional $\ln (\omega)$ from
the upper limit  ($q
\sim 2 k_F$) of the  integration interval.  Thus the total contribution of
this region to
$\rho_{\uparrow \downarrow}$ is of order $T^2 \ln T$.  Notice that this
logarithmic
contribution would be {\it severely suppressed} in the case of the ordinary
Coulomb drag, since the
Fourier transform of the Coulomb interaction between electrons in different
layers
decreases, with increasing $q$, as $e^{-qd}$.

Finally, in region III,   the product $ Im Q_{0 \uparrow}  Im Q_{0
\downarrow}$ is again
proportional to $\omega$, but the integration region in q space is now a
shell of thickness $\sim  \omega/v_F$  centered about $2k_F$.  The
contribution of this
region is therefore of order $T^2$.

Combining the contributions of regions II and III together, after some
calculations, we
arrive at the following expression:
\begin{eqnarray}
&~& \rho_{\uparrow\downarrow}(T)\stackrel{T\ll T_F}{\approx}
 -\frac{\hbar}{e^2}{2\pi^2 r_{s}^2\over  3 (\sqrt{2}+r_{s})^2}
\left({k_B T\over E_F}\right)^2 \cdot \nonumber \\ &~&
\left [   f({r_s\over\sqrt{2}}) -{1\over 2}\ln  \left ({ k_BT\over E_F   }\right ) +\ln 2 +1- {3\over \pi^2}\int_0^\infty 
{y^2\ln y\over \sinh^2 y}dy  \right ],
\label{rhosmallT}
\end{eqnarray}
where $r_{s}=1/\sqrt{\pi n}a^*$ is the dimensionless Wigner-Seitz radius
, $\int_0^\infty 
(y^2\ln y/ \sinh^2 y)dy =-0.55981$
 and the function
$f(x)$ is defined as
 \begin{equation}
f(x) :=  {1\over x-1}\left[1+{x^2+1\over x-1}
\ln ({1+x\over 2x})\right].
\end{equation}
 Thus, {\it in the paramagnetic
case} the spin
trans-resistivity behaves as $T^2 \ln T$ rather than $T^2$.

This is no longer true if the electron
gas is spin-polarized.  Let us assume $n_\uparrow > n_\downarrow$ so that
$k_{F \uparrow} >
k_{F \downarrow}$.  
For a finite polarization and
at sufficiently low
temperatures such that $k_BT\ll E_{F\uparrow}$ and $k_BT\ll E_{F\downarrow}$,
  $Im Q_{0\sigma}(q, \omega)$ can be approximated with its zero-temperature
expression and    
the region of
integration in $q$ coincides with the region in which $Im Q_{0 \downarrow}
(q, \omega)$
differs from zero (see lower panel of Fig.~\ref{fig1}).
  As in the paramagnetic case, it is then possible to divide the
q-integral into  regions in which $Im Q_{0\sigma}$ takes values given by 
Eqs.~(\ref{region1})-(\ref{region3}). 
At the upper limit of  region II for down spins [Eq.~(\ref{region2})], $q =
2k_{F \downarrow} -
\omega /v_{F \downarrow}$,   $Im Q_{0 \downarrow} (q, \omega) $ diverges as
$1/\sqrt{1 -
(q/2k_{F \downarrow})^2}$.  However, $Im Q_{0 \uparrow} (q, \omega) $
remains regular.
Therefore, the logarithmic divergence, leading to the $\ln T$ singularity
in the transresistivity is
cut off.  

Our final result takes the form
\begin{eqnarray}
&~& \rho_{\uparrow\downarrow}(T)\stackrel{T\ll T_F}{\approx}
 -\frac{\hbar}{ e^2}{ \pi^2 r_{s}^2\over  3  (1 - \xi^2)^{3/2}} \left({k_BT\over E_F}\right)^2
g(r_s, \xi)
\label{rhosmallT2}
\end{eqnarray}
 where  
 the
function $g$ is
defined as
\begin{equation}
g(r_s,\xi) = \int_0^1 dx {x \over \left [ x + {r_s \over \sqrt {2 (1 -
\xi)}} \right ]^2 \sqrt {(1 -
\alpha x ) (1 - x) }}~,
\end {equation}
$\xi = (n_\uparrow -
n_\downarrow)/
(n_\uparrow + n_\downarrow)$  is the degree of spin polarization and
$\alpha \equiv \sqrt{1  - \xi} /\sqrt{ 1 + \xi}$.
 Notice that $g
(r_s, \xi)$ diverges
logarithmically for $\xi \to 0$, in agreement with our previous findings.

Fig.~\ref{fig3} focuses on the paramagnetic degenerate regime: it
presents a comparison 
 between the numerical evaluation of $\rho_{\uparrow\downarrow}$ from
Eq.~(\ref{rho})(curve labelled C ) and its analytical approximation
Eq.~(\ref{rhosmallT}) (curve labelled B).
The curve labelled as D is obtained neglecting the $T^2\ln T$ term in
Eq.~(\ref{rhosmallT}): by comparing the D curve with the B curve, the importance of 
the  $T^2\ln T$ correction becomes evident: 
without it the 
spin drag is strongly underestimated.
The A curve  represents 
the result obtained using in Eq.~(\ref{rho})  
the
zero-temperature limit for
$Im Q_{0\sigma}(q,\omega)$. As expected, Fig.~\ref{fig3} shows that
in this range of temperatures,
the analytical approximation (B curve) is very close to such result.

\section{The non-degenerate limit}
In  the non-degenerate limit  $k_BT >> E_F$, the dimensionless chemical
potential $\bar
\mu_\sigma$  of Eq.  (\ref{Q2D}) is given by the classical formula
$\bar \mu_\sigma =  \bar T_\sigma  \ln(\hbar^2 2\pi  n_\sigma/m^* k_BT)$.
Since $- \bar
\mu_\sigma/\bar T_\sigma \to  \infty$  we can replace the function $F(z)$
(Eq. (\ref{defF})) by
its limiting form  $F(z) \to \sqrt{\pi}/2z$ for $z \to \infty$.

This leads to the result
\begin{equation} Im Q_{0\sigma}({q},{\omega})\stackrel{T\gg
T_F}{\approx} {1 \over a^*} {E_{F \sigma} \over k_BT}  \sqrt{{m^* \pi \over
2 k_BT}} {\omega\over q^2}e^{-{  m^*\omega^2\over 2k_BT q^2}}
e^{-  {\hbar^2 q^2\over 8 m^* k_BT}},
\label{Q_high_T}
\end{equation}
identical  to the  3D case.  The Gaussian factor  $e^{-  \hbar^2 q^2\over 8
m^* k_BT}$ on the
right hand side of this equation tends to $1$ in the classical limit (i.e.,
for $\hbar \to 0$) for any
finite wave vector.    However,  we cannot set $\hbar = 0$,  because the
integral
extends to arbitrarily large wave vectors, and the classical approximation
fails at
 large enough values of $q$.  The Gaussian factor assures the convergence
of the integral  in
Eq.~(\ref{rho}) by cutting off the integral at values of $q$ such that $
\hbar^2 q^2/2 m^* \sim
k_BT$.

Using the  classical expression for the dielectric constant $\epsilon(q,\omega)\approx
1+k_D/q$, with $k_D=2\pi e^2 n/\epsilon k_BT$, after simple
 manipulations,
we finally obtain 
\begin{eqnarray}
\rho_{\uparrow\downarrow}(T) &\stackrel{T\gg T_F}{\approx}&
-\frac{\hbar}{e^2}\pi^{3/2}{Ry^*\over k_BT}\int_0^\infty dx{\sqrt{x}e^{-x}
\over(\sqrt{x}+\lambda)^2} \label{rhohighT1}\\
&\stackrel{\lambda\to 0}{\approx}&
-\frac{\hbar}{e^2} \pi^2 {Ry^*\over k_BT},
\label{rhohighT2}\end{eqnarray}
where $\lambda\equiv (2^{3/2}/r_s^2)(Ry^*/K_BT)^{3/2}$ and  
 $Ry^* = m^*e^4/2 \epsilon^2 \hbar^2$ is the effective Rydberg.

Remarkably, the last  result (corresponding to $k_BT\gg Ry^*$ or $r_s\gg 1$)
is {\it independent of the electronic density}.
  It
differs from the
analogous result in three dimensions, which scales as $1/T^{3/2}ln(T)$ and has a
weak (logarithmic)
dependence on the electronic density \cite{long}.

The limiting form of the  transresistivity  Eq.~(\ref{rhohighT2})   is
shown in Fig.~\ref{fig4}
(dashed-dot line)  along with the results obtained from Eq.~(\ref{rho})
for the same choice
of carrier densities as in Fig.~\ref{fig2}.
We stress that this curve delimits
 the region of the $(n,T)$ plane occupied by the family of curves
$\rho_{\uparrow\downarrow}(n,T)$
and  is approached at lower and lower temperatures for decreasing densities
(see Fig.~\ref{fig4}).
The inset of Fig.~\ref{fig4} shows the comparison between
$\rho_{\uparrow\downarrow}$
calculated for $n=10^{10}cm^{-2}$ (solid line)
and its limiting behaviors Eq.~(\ref{rhohighT1}) (dashed line) and  Eq.~(\ref{rhohighT2})(dashed-dot line).
It is interesting to notice that, because of the low
Fermi temperatures and at variance with the 3D case, in  2D the
  non-degenerate behavior of the
spin-transresistivity could be  observed experimentally,
since it corresponds, for reasonable  densities,
to temperatures as low as 100K (see inset  Fig.~\ref{fig4}).
We also notice that the non-degenerate limit
Eq.~(\ref{rhohighT1}) is in reality a good approximation even for
temperatures as low as few $T_F$.
\section {Discussion and summary}
A few comments are now in order: as Eq.~(\ref{rhosmallT})  shows,
  in the degenerate regime,
the  2D spin transresistivity behaves as
 $T^2(A+B\ln T)$, in contrast to
 the 3D case in which  $\rho_{\uparrow\downarrow}\sim T^2$ (see
Ref.~\cite{sdrag}). The additional  logarithmic correction is due to the
fast variation of $Im Q_{0\sigma}(q,\omega)$ for $q\approx 2k_{F\sigma}$,
 so that the high $q$ regime dominates the low-temperature behavior of
$\rho_{\uparrow\downarrow}$.
This correction is also  present, in principle,   in the more familiar
Coulomb drag
transresistivity between two separate electron layers. However, in
that case, due to the presence of the insulating barrier of thickness $d$,
 the Fourier transform of the interlayer Coulomb interaction is given by
$2 \pi
e^2 e^{-qd}/\epsilon q$ and decays exponentially   for large  $q$ .  Thus,
in CD  the influence of
the logarithmic term on the behavior  of $\rho_{\uparrow\downarrow}$ in the
degenerate regime
is negligible.

The exponential decay of the   interlayer interaction is responsible for
another main
difference between the Coulomb drag and the spin Coulomb drag. As can be
seen in
Ref.~\cite{plasmon}, for $T\stackrel{<}{\sim}T_F$ the transresistivity is dominated by the
plasmon modes (especially the acoustic one),
 which enhance the  contribution of small wave vectors. In the
2D-SCD, however, 
the acoustic mode, in which the two electron populations oscillate out of 
phase, is absent.  Fig.~\ref{acoustic} shows how the function
$I(\omega,q)\equiv \exp(-2dq)
 Im Q_{0\uparrow}(q,\omega )
Im Q_{0\downarrow}(q,\omega )/4\left|\epsilon(q,\omega )\right|^2
\sinh^2(\hbar\omega /2k_BT)$,
behaves for small (upper panel) and intermediate (lower panel) values of $q$ in the paramagnetic regime.
For $d=0$, $I(\omega,q)$ corresponds to the integrand of Eq.~(\ref{rho});
in the figure, however, the interaction between the two spin populations is taken as 
$v_{\uparrow\downarrow}=(2\pi e^2/q)\exp(-qd)$ and the RPA
dielectric function as $\epsilon
(q,\omega) = [1 - v(q)\chi_{0 \uparrow} (q,\omega)][1  - v(q)\chi_{0\downarrow}(q,\omega)]
-v_{\uparrow\downarrow}\chi_{0 \uparrow} (q,\omega)\chi_{0\downarrow} 
(q,\omega)$.
 Different curves correspond to different values of $d$ ($0\le d/k_F\le 9$, as labelled).  
Let us first focus  on the upper panel in which $q=0.1k_F$: for $d=0$ the main contribution to the integral comes from the single-pair excitation 
continuum. 
In this case the only possible collective mode (corresponding to  the zeros of $\epsilon
(q,\omega)$)
is the in-phase optical mode, whose contribution (the spike labelled as $0$) is in any 
case negligible.
As the exponential factor in the interaction $v_{\uparrow\downarrow}$
is turned on 
(i.e. as the scattering rate between the two spin population is decreased),
 the strength of the 
integrand is transfered from the single-pair excitation continuum to the 
collective modes and the out-of-phase acoustic 
plasmon appears (see the two maxima labeled with  $9$, corresponding to
$d=9k_F$).

The lower panel of Fig.~\ref{acoustic} 
presents $I(\omega,q)$ for $q=0.5k_F$: already at this 
intermediate value, the contribution to $\rho_{\uparrow\downarrow}$ 
 becomes negligible when 
$d>0$: in fact $I(\omega,q)$  scales as 
$v_{\uparrow\downarrow}^2\sim\exp(-2qd)$.  

We would like to emphasize  a more general point: as 
Fig.~\ref{acoustic} clearly shows, due to the absence of the exponential factor
$e^{-qd}$ in the Fourier transform of the Coulomb potential, 
the drag effect in the SCD
is definitely larger than in the ordinary CD. In other words, in SCD, 
 the two electron populations can transfer momentum 
 to one another 
more effectively.

The plasmon-enhancement is absent in 3D as well. In this case,
however, this
is due to a combination of three different effects:
 (i) the 3D plasmon frequency $\omega_p(T)$ is finite
 at any temperature ($\omega_p(T)\ge \sqrt{16\pi na^{*3}}Ry^*/\hbar$),
(ii) due to the $\sinh^2(\hbar\omega/2k_BT)$ term in the denominator,
the integrand in Eq.~(\ref{rho}) is significantly large only for
$\hbar\omega\stackrel{<}{\sim} k_BT$, while it decreases exponentially 
for higher temperatures, and, (iii)
 the plasmon line-width,  increases with temperature.
Points (i) and (ii) imply that for small temperatures the integrand
has already vanished when  $\omega\approx\omega_p(T)$; on the other side,
because of point (iii),
for temperatures such that $k_BT \stackrel{<}{\sim}\omega_p(T)$, 
the strength of the
plasmon is negligible.

Let us now examine some  issues concerning the experimental observation of the SCD.

First of all, it must be clear  that the SCD is an intrinsic many-body effect, and, therefore,  it is possible to design an experiment  to measure it directly  and independently of the ordinary diagonal part of the resistivity tensor. Such an experiment has been described in Refs. \cite{sdrag,long}.
    
For different experimental setups, however,  it might be important  to know how the spin drag resistivity compares to the familiar Drude resistivity $\rho_D$.  This information is provided in Fig.~\ref{fig6}
for doped layers  of  InAs ($m^*=0.026m_e$, $\epsilon=13.6$)  and GaAs ($m^*=0.067m_e$, $\epsilon=12$) at two different densities.   
Since it is customary to express the resistivity of doped semiconductor samples in terms of their mobility,  in the upper panel 
we plot the quantity $\mu_{\uparrow \downarrow} = - 1/n e \rho_{\uparrow \downarrow}$ as a function of temperature. $\mu_{\uparrow \downarrow}$ has the dimensions of a mobility, and its value should be compared to that of the ordinary mobility $\mu_D$ of the sample :  if, at a given temperature,  $\mu_{\uparrow \downarrow} < \mu_D$ the spin drag resistivity is larger than the ordinary Drude resistivity. 
It is evident from this figure that for appropriate but realistic parameters, $\mu_{\uparrow\downarrow}$  and $\mu_D$ can be quite comparable. 
The significant numerical difference between the  $InAs$ and $GaAs$  results is  primarily due to the larger effective mass, and, to a lesser extent, to the smaller dielectric constant of $GaAs$.  
The impact of these two parameters on the spin drag resistivity is easily understood.  Keeping in mind  that the SCD is a consequence of Coulomb 
scattering between different spin populations,
a smaller dielectric constant means stronger Coulomb scattering, and a larger effective mass implies a higher density of states available for scattering. 
Notice that the minimum value of  $\mu _{\uparrow\downarrow}$ decreases with
increasing density.

In the lower panel of Fig.~\ref{fig6}, we plot
the ratio $\rho_{\uparrow\downarrow}/\rho_D$ (which is 
proportional to the sample
mobility) versus  temperature 
 for a 
mobility $\mu_D = 3\times 10^{3} cm^2/Vs$.   
The same parameters and materials of the upper panel have been chosen.
As the figure shows, for reasonable temperatures and realistic parameters,
 $\rho_{\uparrow\downarrow}$ can be a large fraction of $\rho_D$.

A peculiar signature of the  two-dimensional SCD,
which  would be interesting to  observe, 
is the  non-degenerate  density-independent behavior given by Eq.(\ref{rhohighT2}). 
As shown in the inset of Fig.~\ref{fig4}, a sample
would display such a behavior at $T>>T_F$, 
i.e. in the high temperature/low density\cite{lowdenRPA} 
regime. Since, for densities
of the order of $n=10^{10}cm^{-2}$, $T_F$ is as low as $4K$ in GaAs  and $11K$ in InAs, this regime should be observable in these materials for temperature considerably lower than room temperature.

Our calculations  did not include  the finite width of a realistic sample.   The response of a  quasi-2D sample  would be somewhere in between our 3D and  2D results, depending on how well the idealized two dimensional approximation is fulfilled. 
In both the 2D and 3D cases however, we see that the ratio 
$\rho_{\uparrow\downarrow}/\rho_D$     
remains of the same order of magnitude for the same material and mobility and  comparable density(see Fig.~\ref{fig6}, lower panel, and  Ref.~\cite{long}, Fig.~2).
Moreover even in the fully three dimensional case, the non-degenerate regime is only weakly (logarithmically) dependent on
the carrier density. The above observations imply that
 a modest finite size correction would not modify 
significantly the relative importance of the predicted effect, or its non-degenerate behavior.   
 
In summary we have discussed the effect of dimensionality on the
spin-transresistivity and shown that in two, as well as in
three\cite{EPL,long} dimensions,
the  spin Coulomb drag can be a sizeable effect. We have discussed the
differences
between   Coulomb drag and spin Coulomb drag effects showing how,  in the
degenerate
regime, the  different form of the Fourier transform of the Coulomb interaction
determines a
quantitatively different
 behavior of the spin-transresistivity in comparison to the
CD-transresistivity. We have also explained why in general the spin Coulomb
drag is expected to be larger than the ordinary Coulomb drag.
Finally we have demonstrated that, in the non-degenerate two-dimensional
regime, the spin-transresistivity displays a universal density-independent
behavior.
Such a behavior could be in principle observed experimentally for realistic
system
parameters.

\section {Acknowledgements} We gratefully acknowledge support from NSF grant No. DMR-0074959.

\begin{figure}
  \caption{Spin transresistivity $\rho_{\uparrow\downarrow}$ as a function
of temperature (rescaled by $T_F$) for InAs parameters ($m^*=0.026m_e$,
$\epsilon=13.6$). Each curve corresponds to a
different density as labeled.   }
  \label{fig2} \end{figure}

\begin{figure}
  \caption{Parameter space for the existence of a non-zero
  $Im Q_{0\sigma}(q,\omega)$ at $T=0$.
\\ Upper panel: paramagnetic case ($k_{F\uparrow}/k_{F\downarrow}=1$).
The three curves correspond to the set of points in which one or both of
the arguments of the Heaviside functions in Eq.~(\ref{chi_T0}) are
equal to zero: $(q/k_F)^2/2+ q/k_F$ (solid curve), $-(q/k_F)^2/2+ q/k_F$
(dashed curve) and $(q/k_F)^2/2- q/k_F$ (dotted curve). $Im Q_{0\sigma}(q,\omega)$
 is different from zero
 in the region between the solid and the dotted curve. The points
labeled by A to D correspond respectively to 
$q_{A\sigma}(\omega)$ to $q_{D\sigma}(\omega)$ for $\hbar\omega/2E_{F\sigma}
 = 0.2$.
\\ Lower panel: spin-polarized case ($k_{F\uparrow}/k_{F\downarrow}=1.5$).
 The solid curve define the parameter space corresponding to   
$ImQ_{0\uparrow}(q,\omega)$, the dashed curve the one related to
$ImQ_{0\downarrow}(q,\omega)$.  The points
labeled as B$\downarrow$ and C$\downarrow$ correspond to $q_{B\downarrow}$ and
$q_{C\downarrow}$ respectively, and delimit the interval relevant to the calculation of Eq.~(\ref{rhosmallT2}).}
  \label{fig1}
\end{figure}

\begin{figure}
  \caption{Degenerate regime: Comparison between
$\rho_{\uparrow\downarrow}$ (curve labelled C), its analytical approximation
Eq.~(\ref{rhosmallT})
(curve labelled B), the approximation obtained using the zero-temperature form
of
$Im Q_{0\sigma}(q,\omega)$
in
 Eq.~(\ref{rho})
 (curve labelled A), and the analytic one obtained by neglecting the $T^2\ln T$ term 
in Eq.~(\ref{rhosmallT})(curve labelled D). The curves are plotted
  vs temperature (rescaled by $T_F$) for $n=10^{10} cm^{-2}$ and InAs parameters.
 }
  \label{fig3}
\end{figure}

\begin{figure}
  \caption{Non-degenerate regime: Comparison between
 $\rho_{\uparrow\downarrow}$ (solid line) and its universal limiting behavior
(dash-dotted line)
 Eq.~(\ref{rhohighT2}) vs temperature for different carrier densities and InAs parameters.
  Inset: Comparison between
$\rho_{\uparrow\downarrow}$ (solid line) and  its  approximations
 (Eq.~(\ref{rhohighT1}) (dashed line)
 and  Eq.~(\ref{rhohighT2})
(dashed-dotted line)). The curves are calculated for $n=10^{10} cm^{-2}$ and plotted
 vs temperature (rescaled by the Fermi temperature).
   }
  \label{fig4}
\end{figure}

\begin{figure}
  \caption{Behavior of $I(\omega,q)\equiv \exp(-2dq)
 Im Q_{0\uparrow}(q,\omega )
Im Q_{0\downarrow}(q,\omega )/4\left|\epsilon(q,\omega )\right|^2
\sinh^2(\hbar\omega /2k_BT)$ vs the rescaled frequency $\hbar\omega/2E_F$
for fixed $q$ and temperature but different values of $d$ ($0\le d/k_F\le 9$).\\
Upper panel: small-$q$ behavior ($q=0.1k_F$). 
Each curve corresponds to a different  value of $d/k_F$ as labelled. \\
Lower panel: intermediate-$q$ behavior ($q=0.5k_F$). Curves are  labelled with
the corresponding value of $d/k_F$  (as in the upper panel).
 }
  \label{acoustic}
\end{figure}

\begin{figure}
  \caption{Upper panel: Spin drag mobility $\mu_{\uparrow\downarrow}\equiv
-1/\rho_{\uparrow\downarrow}ne$ vs temperature (rescaled by $T_F$)
for $n=10^{11}cm^{-2}$ (dashed curve) and $n=10^{12}cm^{-2}$ (solid curve).
 Each couple of curves corresponds to a
different material:  InAs ($m^*=0.026m_e$,
$\epsilon=13.6$) and GaAs ($m^*=0.067m_e$,
$\epsilon=12$), as labelled.\\
Lower panel: Ratio $\rho_{\uparrow\downarrow}/\rho_D$ as a function
of temperature for $n=10^{11}cm^{-2}$ (dashed curve) and
 $n=10^{12}cm^{-2}$ (solid curve) and sample
mobility $\mu=3\times 10^3 cm^2/Vs$.  Each couple of curves corresponds to a
different material  (InAs  and GaAs , as labelled)}
  \label{fig6} \end{figure}

\newpage
\begin{figure}
  \psfig{figure=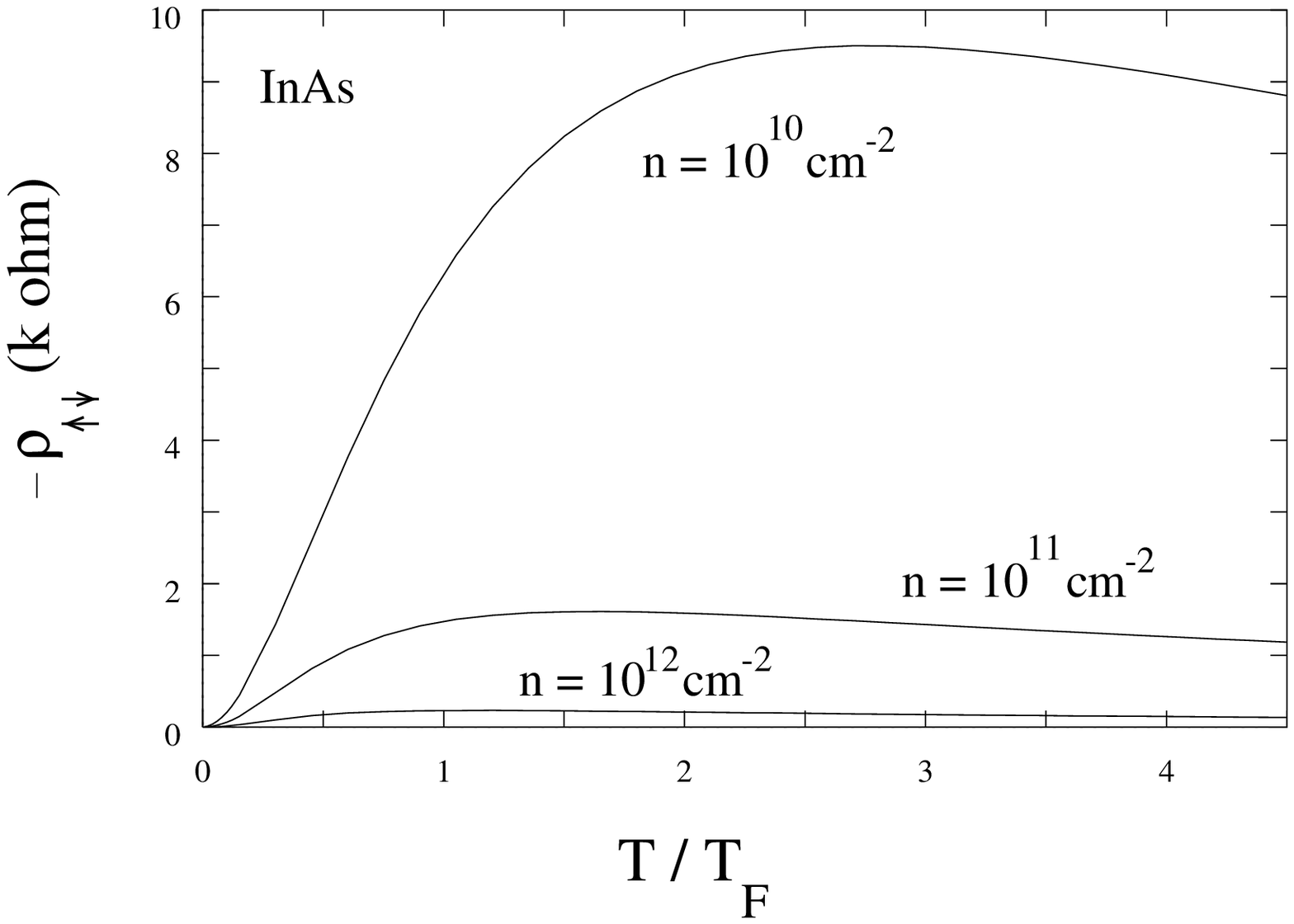,width=1.05\columnwidth,angle=0}
\end{figure}

\newpage
\begin{figure}
  \psfig{figure=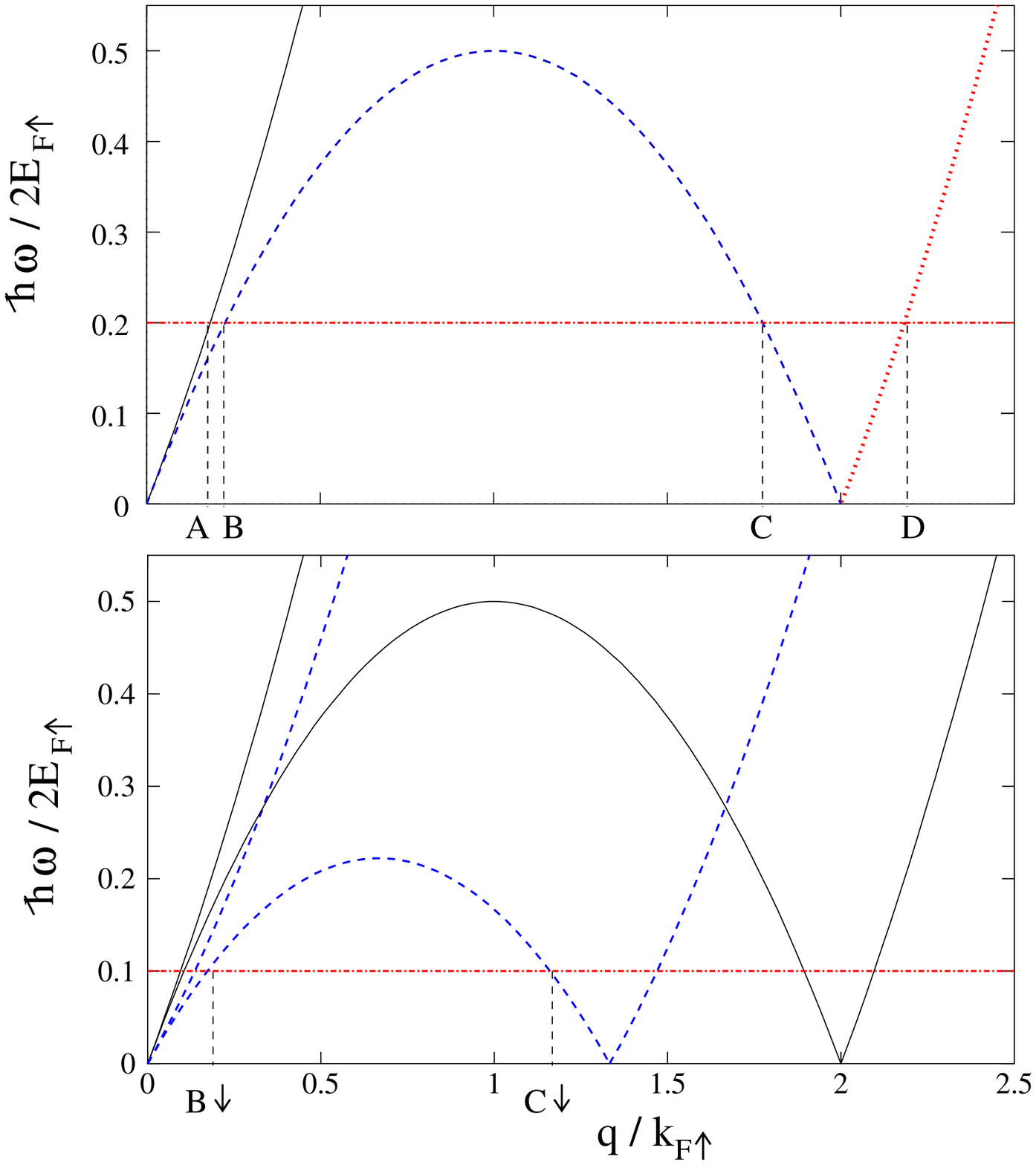,width=1.05\columnwidth,angle=0}
\end{figure}

\newpage
\begin{figure}
  \psfig{figure=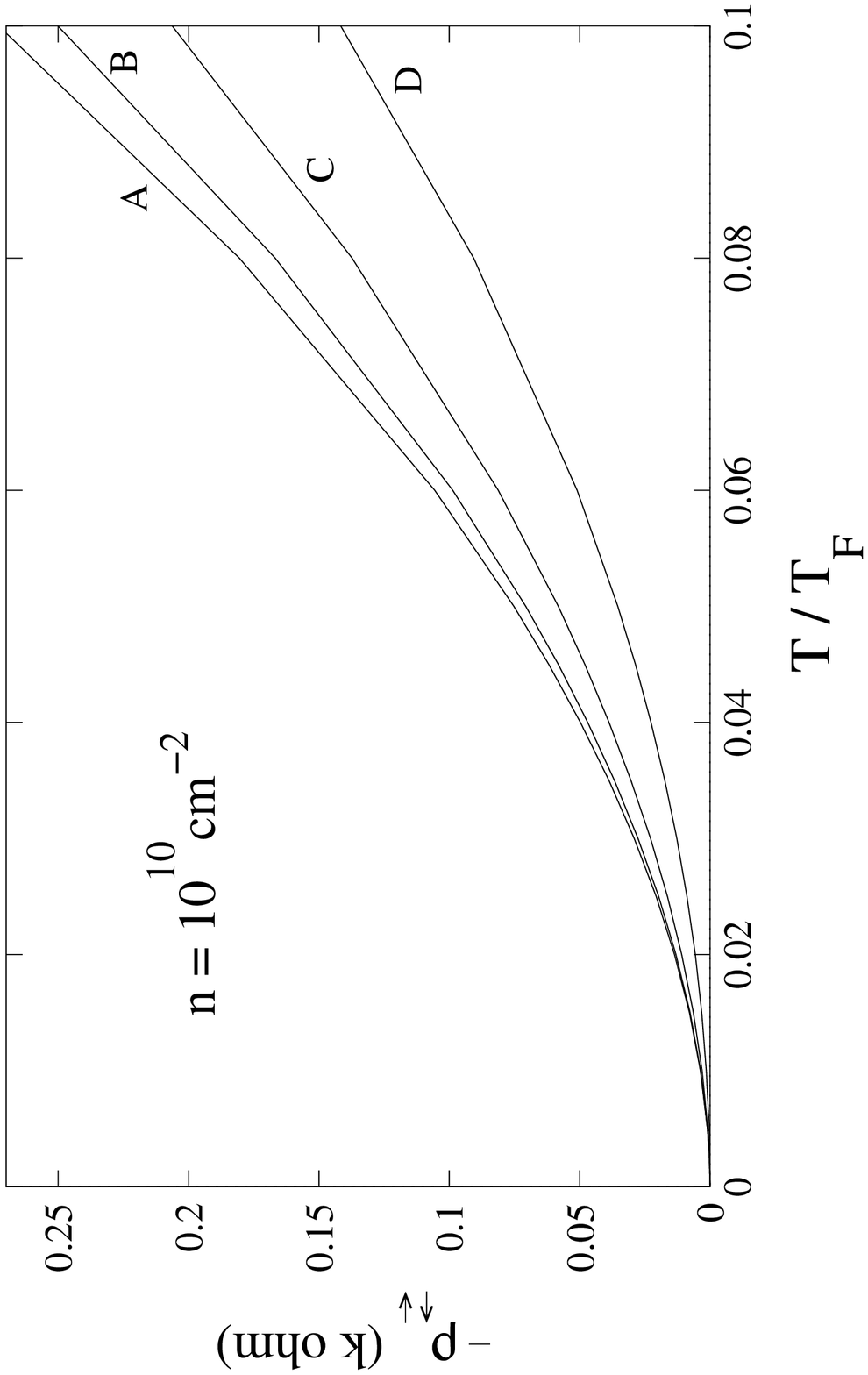,width=1.05\columnwidth,angle=0}
\end{figure}

\newpage
\begin{figure}
  \psfig{figure=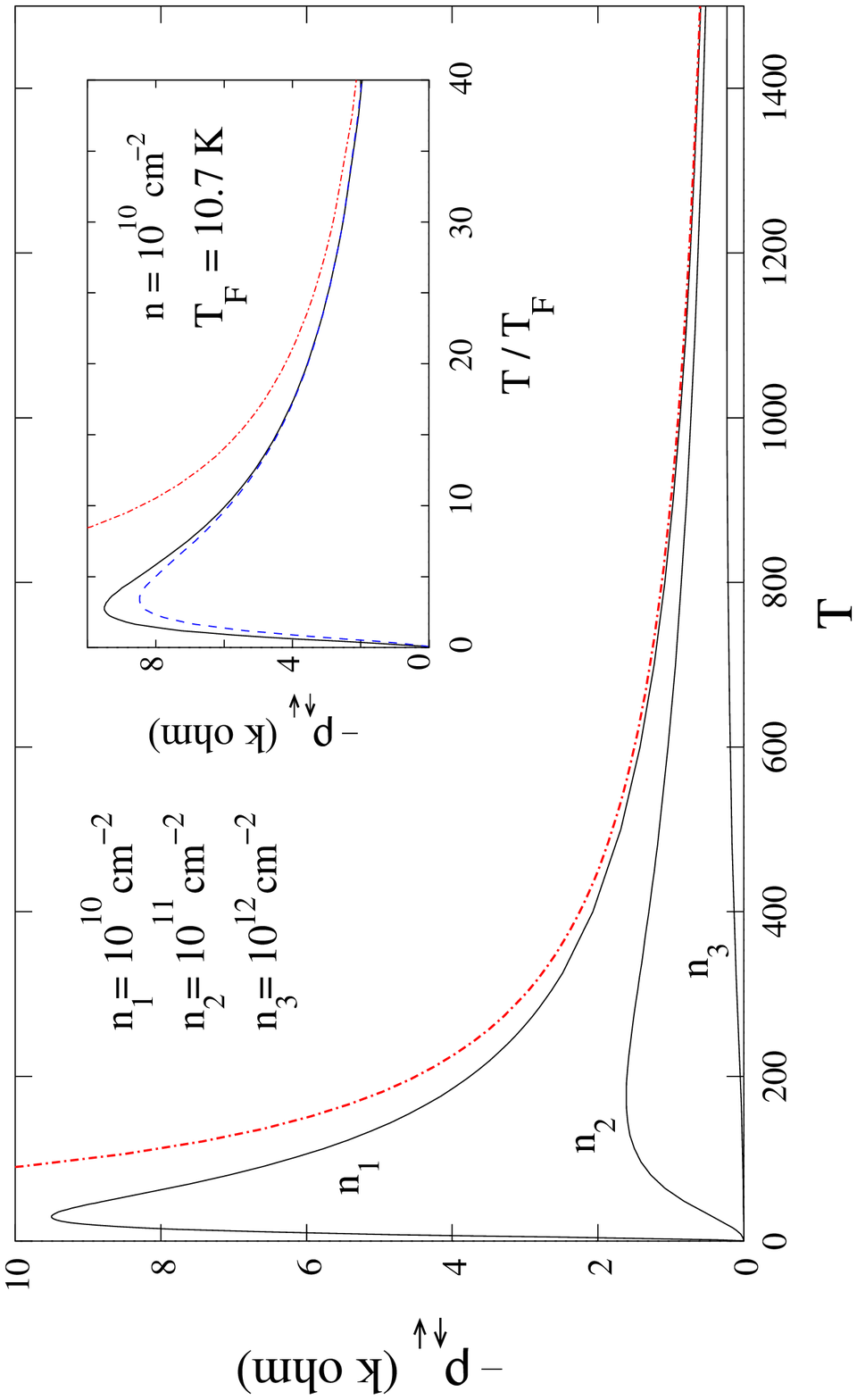,width=1.05\columnwidth,angle=0}
\end{figure}

\newpage
\begin{figure}
  \psfig{figure=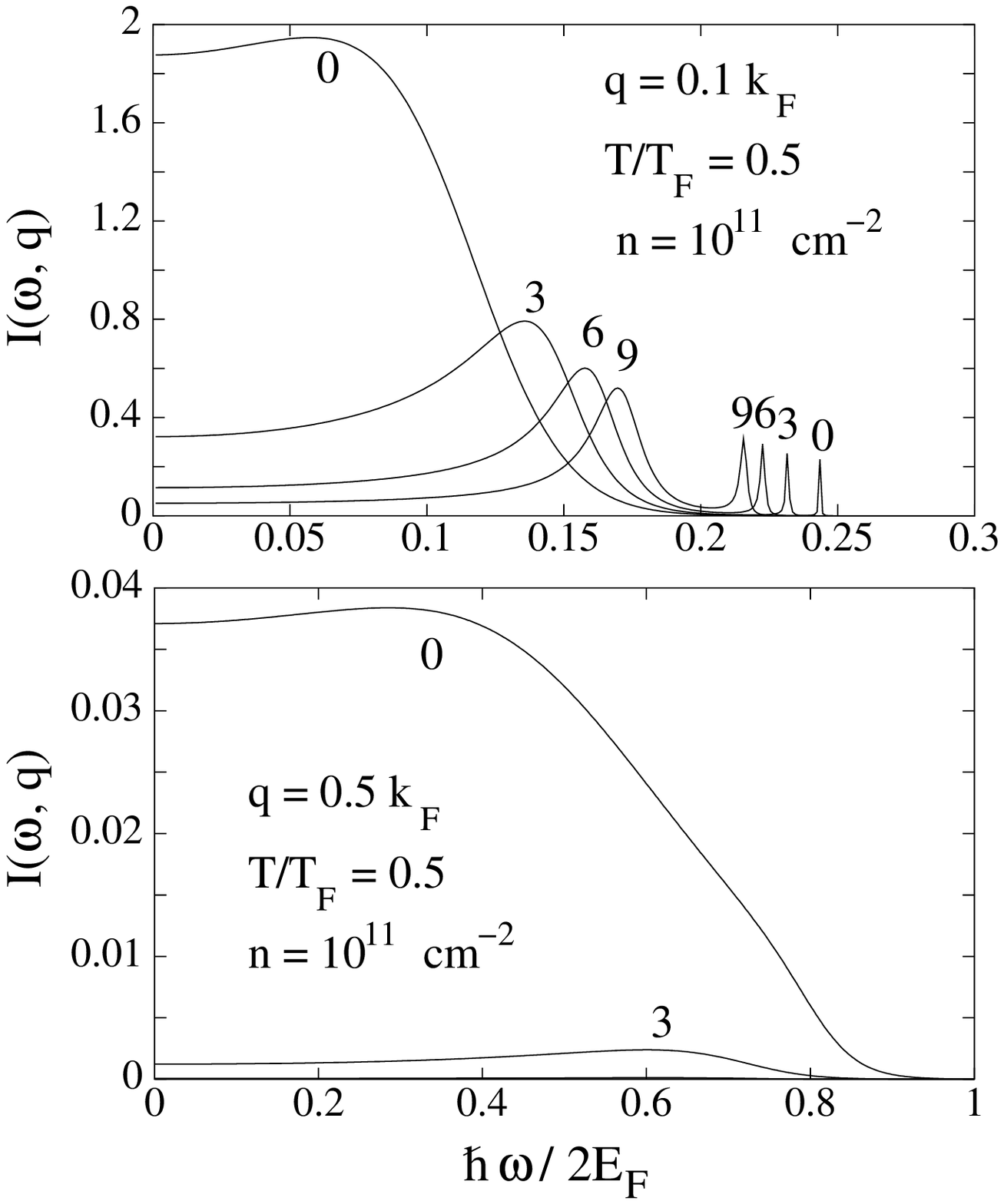,width=1.05\columnwidth,angle=0}
\end{figure}

\newpage
\begin{figure}
  \psfig{figure=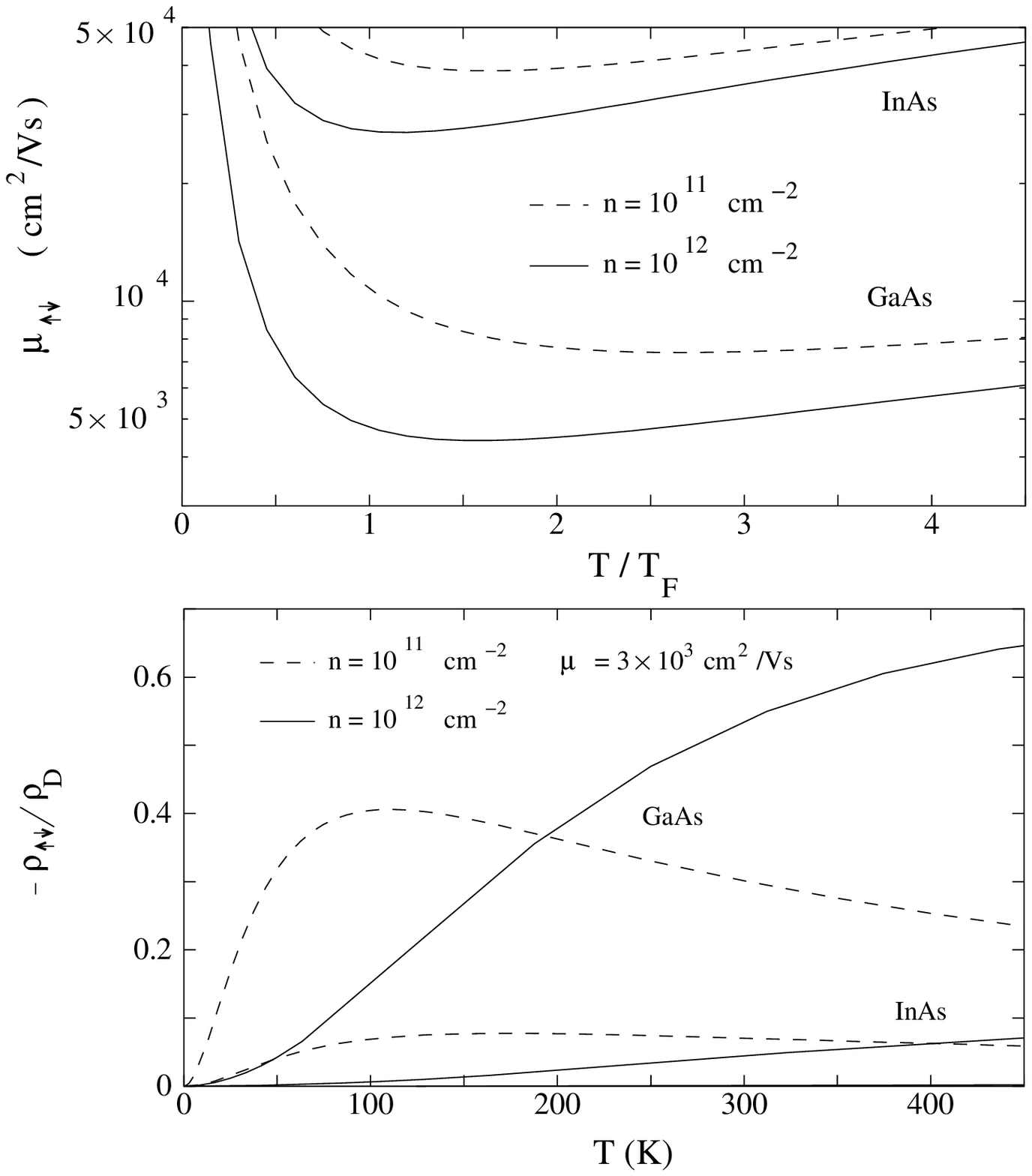,width=1.2\columnwidth,angle=0}
\end{figure}


\begin{thebibliography}{aaaaaaaaaaaaaaaaaaaa}
\bibitem{gen} 
For a recent review on spin-polarized transport related issues,
see S. A.  Wolf et al., {\it Science}  {\bf 294}, 1488 (2001) and references 
therein.
\bibitem{Schmidt}G. Schmidt et al., {\it Phys. Rev. B} {\bf 62}, R4790 (2000).
\bibitem{Ohno} H. Ohno, {\it Science} {\bf 281}, 951 (1998).
\bibitem{sdrag} I.
D'Amico and G. Vignale, {\it Phys. Rev. B} {\bf 62}, 4853 (2000).

\bibitem{EPL}I. D'Amico and G. Vignale,
{\it Europhys. Lett.} {\bf 55}, 566 (2001).
\bibitem{long}I. D'Amico and G. Vignale,  {\it Phys. Rev. B}
{\bf 65}, 085109 (2002)
\bibitem{spin-flip}The issue of spin-flip scattering  has been discussed in detail in \cite{long}. There we argued that this contribution is probably very 
small due
to the short range of the spin-flip scattering potential and the weakness
of that potential. Here we focus on the Coulomb contribution, which we
expect to be larger at experimentally relevant temperatures, and which
contains the interesting physics of electronic correlations.
\bibitem{Rojo} For a recent review of the theoretical and experimental
situation on Coulomb drag see A. G. Rojo, {\it J. Phys.: Cond. Mat} {\bf
11} R31 (1999) and references therein.


\bibitem{impurities}The effect of impurity
correlation on disorder
averaging was examined in  \cite{sdrag}, and shown there to vanish in the
limit of weak
disorder/weak interaction


\bibitem{phonons} In principle the dielectric ``constant''
 should be  frequency
 dependent due to the vibrational modes 
of the crystal. In the present treatment however,
the effect of phonons is incorporated only by choosing as dielectric constant
$\epsilon$
 the average between the static ($\epsilon_0$) and the ``high frequency'' 
($\epsilon_\infty$) dielectric
constants.    

\bibitem{plasmon}K. Flensberg and B. Yu-Kuang Hu, {\it Phys. Rev. Lett.}
{\bf 73 }, 3572 (1994); K. Flensberg and B. Yu-Kuang Hu,  {\it Phys. Rev.
B} {\bf 52 } 14796 (1995);  N.P.R. Hill et al. {\it Phys. Rev. Lett.} {\bf
78}, 2204 (1997);
 H. Noh et al., {\it Phys. Rev. B} {\bf 58 }, 12621 (1998).
\bibitem{Flensberg} An independent  calculation of the 2-dimensional spin
transresistivity has recently been done  by K. Flensberg, T. S. Jensen
and N. A. Mortensen, cond-mat/0107149

\bibitem{footnote} The formal lowest order in the strength of the Coulomb interaction, namely, the second, diverges, due to the infinite range of the interaction. This is remedied by screening the interaction through the inclusion of an infinite series of RPA diagrams, which is expected to be exact in the high density/weak coupling limit. \\ RPA should in any case be 
qualitatively correct at all densities as long as the Landau Fermi
 liquid description of the system is valid.  As a matter of fact,
 experiments on the ordinary Coulomb drag  (see  \cite{Rojo}) 
 were found to qualitatively confirm the main predictions of the RPA even in 
the low-density regime,
 which reinforces our expectations that RPA predictions on the spin Coulomb
 drag will also be qualitatively correct.


\end{thebibliography}
\end{document}